\documentclass[doublecol]{epl2}
\usepackage{amsmath}
\usepackage{amssymb}
\usepackage{graphicx}

\title{High magnetic fields thermodynamics of heavy fermion metal $\rm \bf YbRh_2Si_2$.}
\shorttitle{High magnetic fields thermodynamics}
\author{V. R. Shaginyan \inst{1}\thanks {Email:
\email{vrshag@thd.pnpi.spb.ru}} \and K. G. Popov\inst{2}\and V. A.
Stephanovich\inst{3}\thanks {Email: \email{stef@math.uni.opole.pl}
and Homepage: http://draco.uni.opole.pl/
$\sim$stefan/VStephanovichDossier.html} \and V. I. Fomichev\inst{4}
\and E. V.Kirichenko\inst{5}} \shortauthor{V. R. Shaginyan \etal}

\institute{\inst{1} Petersburg Nuclear
Physics Institute, Gatchina, 188300, Russia\\
\inst{2} Komi Science Center, Ural Division, RAS, Syktyvkar,
167982, Russia\\ \inst{3} Opole University, Institute of Physics,
Opole, 45-052, Poland\\ \inst{4} St. Petersburg Division of the
Classical Academy, Varshavskaya street 50/2, St. Petersburg,
Russia\\ \inst{5} Opole University, Institute of Mathematics and
Informatics, Opole, 45-052, Poland}

\abstract{We perform comprehensive theoretical analysis of high
magnetic field behavior of the heavy-fermion (HF) compound $\rm
YbRh_2Si_2$. At low magnetic fields $B$, $\rm YbRh_2Si_2$ has a
quantum critical point related to the suppression of
antiferromagnetic ordering at a critical magnetic field $B\perp c$
of $B=B_{c0}\simeq0.06$ T. Our calculations of the thermodynamic
properties of $\rm YbRh_2Si_2$ in wide magnetic field range from
$B_{c0}\simeq0.06$ T to $B \simeq 18$ T allow us to straddle a
possible metamagnetic transition region and probe the properties of
both low-field HF liquid and high-field fully polarized one.
Namely, high magnetic fields $B\sim B^*\sim 10$ T fully polarize
corresponding quasiparticle band generating Landau Fermi liquid
(LFL) state and suppressing HF (actually NFL) one, while at
elevating temperatures both HF state and corresponding NFL
properties are restored. Our calculations are in good agreement
with experimental facts and show that the fermion condensation
quantum phase transition is indeed responsible for the observed NFL
behavior and quasiparticles survive both high temperatures and high
magnetic fields.}

\begin{document}
\maketitle

\begin{figure} [! ht]
\begin{center}
\vspace*{-0.5cm}
\includegraphics [width=0.47\textwidth]{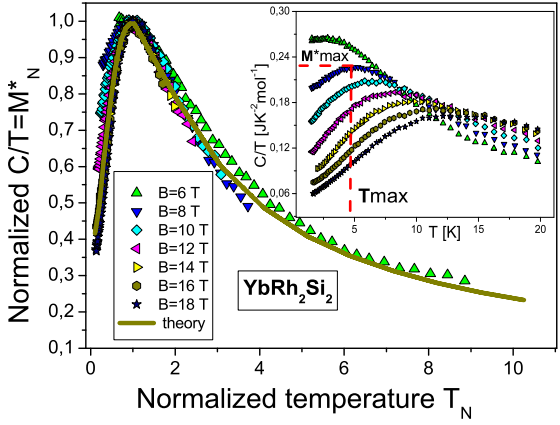}
\vspace*{-0.8cm}
\end{center}
\caption{The normalized effective mass $M^*_N=M^*/M^*_{\rm max}$
versus normalized temperature $T_N=T/T_{\rm max}$. $M^*_N$ is
extracted from the measurements (shown in the inset) of the
specific heat $C/T$ on $\rm YbRh_2Si_2$ in magnetic fields $B$
\cite{steg1} listed in the legend. Our calculations (made at
$B\simeq B^*$ when the quasiparticle band is fully polarized) are
depicted by the solid curve tracing the scaling behavior of
$M^*_N$. The inset reports the temperature dependence of the
electronic specific heat $C/T$ of $\rm YbRh_2Si_2$ at different
magnetic fields \cite{steg1} shown in the main panel legend. The
illustrative values of $M^*_{\rm max}$ and $T_{\rm max}$ at $B=8$ T
are also shown.}\label{MRM}
\end{figure}

An explanation of the rich and striking behavior of heavy fermion
(HF) metals is, as years before, among the main problems of modern
condensed matter physics. One of the most interesting and puzzling
issues in the research of HF compounds is their non-Fermi liquid
(NFL) behavior in a wide range of temperatures $T$ and magnetic
fields $B$. For example, recent measurements of the specific heat
$C$ of $\rm YbRh_2Si_2$ under the application of magnetic field $B$
show that the above temperature range extends at least up to twenty
Kelvins as reported in the inset to fig. \ref{MRM}. As it is
well-known from Landau Fermi liquid (LFL) theory, the ratio $C/T$
is proportional to quasiparticle effective mass $M^*$. The inset to
fig. \ref{MRM} reports the dependence of $C(T)/T$, which has a
maximum $M^*_{\rm max}(B)$ at some temperature $T_{\rm max}(B)$. It
is seen from the inset, that $M^*_{\rm max}(B)$ decreases as
magnetic field $B$ grows, while $T_{\rm max}(B)$ shifts to higher
$T$ reaching $15$ K at $B=18$ T \cite{steg1}.

A deeper insight into the behavior of $C/T$ in the inset to fig.
\ref{MRM} can be achieved using some "internal" scales. Namely,
near QCP it is convenient to divide the effective mass $M^*$ and
temperature $T$ by their maximal values, $M^*_{\rm max}$ and
$T_{\rm max}$ respectively. This generates the normalized effective
mass $M^*_N=M^*/M^*_{\rm max}$ and temperature $T_N=T/T_{\rm max}$
\cite{prep}. In the main panel of fig. \ref{MRM} the obtained
dependence $M^*_N(T_N)$ is shown by symbols, corresponding to
different magnetic fields. This immediately reveals the scaling in
the normalized experimental curves - the curves at different
magnetic fields $B$ merge into a single one in terms of the
normalized variable $T_N=T/T_{\rm max}$. It is seen from fig.
\ref{MRM}, that the normalized effective mass $M^*_N(T_N)$ is not a
constant as it would be for LFL case. Rather, it shows the scaling
behavior in normalized temperature $T_N$. It is also seen from fig.
\ref{MRM} (both the main panel and inset) that the NFL behavior and
the associated scaling extend at least to temperatures up to twenty
Kelvins.

Thus, we conclude that a challenging problem for theories
considering the high magnetic field ( $B\sim B^*$) NFL behavior of
the HF metals is to explain both the scaling and the shape of
$M^*_N(T_N)$. Another part of the problem is the remarkably large
temperature and magnetic field ranges where the NFL behavior and
scaling are observed.

In this letter, based on the theory of fermion condensation quantum
phase transition (FCQPT) \cite{prep} we analyze the thermodynamic
properties of $\rm YbRh_2Si_2$ at both low and high magnetic
fields. Our calculations of the specific heat and magnetization
allow us to conclude that under the application of magnetic field
the heavy-electron system of $\rm YbRh_2Si_2$ evolves continuously
without a metamagnetic transition. At low temperatures and high
magnetic fields $B\simeq B^*$ the system is completely polarized
and demonstrates the LFL behavior, while at elevated temperatures
the HF behavior and related NFL one are restored. The obtained
results are in good agreement with experimental facts in the entire
magnetic field ($0.1$ T - $18$ T) and temperature (40 mK - 20 K)
domains.

In our FCQPT approach \cite{prep}, to study the (generally speaking
NFL) behavior of the effective mass $M^*(T,B)$, we simply use
Landau equation for the quasiparticle effective mass in a Fermi
liquid. The only modification is that in our formalism the
effective mass is no more constant but depends on temperature,
magnetic field and other external parameters. For the model of
homogeneous HF liquid at finite temperatures and magnetic fields,
this equation acquires the form \cite{land,prep}
\begin{eqnarray}
\nonumber \frac{1}{M^*_{\sigma}(T,
B)}&=&\frac{1}{m}+\sum_{\sigma_1}\int\frac{{\bf p}_F{\bf
p}}{p_F^3}F_
{\sigma,\sigma_1}({\bf p_F},{\bf p}) \\
&\times&\frac{\partial n_{\sigma_1} ({\bf
p},T,B)}{\partial{p}}\frac{d{\bf p}}{(2\pi)^3}, \label{HC1}
\end{eqnarray}
where $m$ is a bare electron mass, $F_{\sigma,\sigma_1}({\bf
p_F},{\bf p})$ is the Landau amplitude, which depends on Fermi
momentum $p_F$, momentum $p$ and spin $\sigma$. Here we use the
units where $\hbar=k_B=1$. For definiteness, we assume that the HF
liquid is 3D liquid. The Landau amplitude has the form \cite{land}
\begin{equation}\label{lampl}
F_{\sigma,\sigma'}({\bf p},{\bf p'})= \frac{\delta^2 E[n]}{\delta
n_{\sigma}({\bf p})\delta n_{\sigma'}({\bf p'})},
\end{equation}
where $E[n]$ is the system energy, which is a functional of the
quasiparticle distribution function $n$\cite{prep,land}. It can be
expressed as
\begin{equation}
n_{\sigma}({\bf p},T)=\left\{ 1+\exp
\left[\frac{(\varepsilon({\bf p},T)-\mu_{\sigma})}T\right]\right\}
^{-1},\label{HC2}
\end{equation}
where $\varepsilon({\bf p},T)$ is the single-particle spectrum. In
our case, the chemical potential $\mu$ depends on the spin due
to Zeeman splitting $\mu_{\sigma}=\mu\pm \mu_BB$, $\mu_B$ is
Bohr magneton.

In LFL theory, the single-particle spectrum is a variational
derivative of the system energy $E[n_{\sigma}({\bf p},T)]$ with
respect to occupation number $n$, $\varepsilon({\bf p},T)=\delta
E[n({\bf p})]/\delta n$. Choice of the amplitude is dictated by the
fact that the system has to be at the quantum critical point (QCP)
of FCQPT. Namely, in this region the momentum-dependent part of
Landau amplitude can be taken in the form of truncated power series
$F= a({\bf p}-{\bf p}')^2+ b({\bf p}-{\bf p}')^3+c({\bf p}-{\bf
p}')^4+...$, where $a,b$ and $c$ are fitting parameters. We note
that this interaction, being an analytical function of $({\bf
p}-{\bf p}')^2$, can generate topological phase transitions
interfering in FCQPT \cite{prep}. In our case $F$ does not depend
on the number density $x$ of the system as it is fixed by condition
that the system is situated in QCP of FCQPT. Thus, the variational
procedure, being applied to the functional $E[n_{\sigma}({\bf
p},T)]$, gives following form for $\varepsilon({\bf
p},T)=\varepsilon_\sigma({\bf p},T)\equiv
\varepsilon[n_{\sigma}({\bf p},T)]$
\begin{equation}\label{epta}
\varepsilon_\sigma({\bf p},T)=\frac{p^2}{2m}+\sum_{\sigma_1}\int F_
{\sigma,\sigma_1}({\bf p},{\bf p}_1)n_{\sigma_1}({\bf
p}_1,T)\frac{d^3p_1}{(2\pi)^3}.
\end{equation}
Equations \eqref{HC2} and \eqref{epta} constitute the closed set
for self-consistent determination of $\varepsilon_\sigma({\bf
p},T)$ and $n_{\sigma}({\bf p},T)$. The solution of eq.
\eqref{epta} generates the spectrum where the first two
$p$-derivatives equal zero. Since the first derivative is
proportional to the reciprocal quasiparticle effective mass
$1/M^*$, its zero just signifies QCP of FCQPT. The second
derivative must vanish also. Otherwise $\varepsilon(p)-\mu$ has the
same sign below and above the Fermi surface, and the Landau state
becomes unstable \cite{khodb,prep}. Zeros of these two subsequent
derivatives mean that the spectrum $\varepsilon({\bf p})$ has an
inflection point at $p_F$ so that the lowest term of its Taylor
expansion is proportional to $(p-p_F)^3$. In other words, close to
FCQPT the single - particle spectrum does not have customary form
$v_F(p-p_F)$, $v_F$ is fermi velocity.

Having solved eqs. \eqref{HC2} and \eqref{epta}, we substitute
their solution into eq. \eqref{HC1} to obtain field and temperature
dependence of Landau quasiparticle effective mass. We emphasize
here, that in our approach the entire temperature and magnetic
field dependence of the effective mass is brought to us by
dependencies of $\varepsilon_\sigma({\bf p},T)$ and
$n_{\sigma}({\bf p},T)$. The sole role of Landau amplitude is to
bring the system to FCQPT point, where Fermi surface alters its
topology so that the effective mass acquires temperature and field
dependence, see Ref.\cite{prep} and references therein for details.

Rewriting the quasiparticle distribution function as
$n_{\sigma}({\bf p},T,B) \equiv n_{\sigma}({\bf p},T=0,B=0)+\delta
n_{\sigma}({\bf p},T,B)$ yields more convenient form for the
equation (\ref{HC1})
\begin{eqnarray}
\nonumber
&&\frac{1}{M^*(T,B)}=\frac{1}{M^*}+\frac{1}{p_F^2}\sum_{\sigma_1}
\int\frac{{\bf p}_F{\bf p_1}}{p_F}\\
&\times&F_{\sigma,\sigma_1}({\bf p_F},{\bf
p}_1)\frac{\partial\delta n_{\sigma_1}({\bf p}_1,T,B)}
{\partial{p}_1}\frac{d{\bf p}_1}{(2\pi) ^3}. \label{HC3}
\end{eqnarray}

Our analysis shows, that near FCQPT the normalized solution of eq.
\eqref{HC3} $M^*_N(y=T_N)$ can be well approximated by a simple
universal interpolating function. The interpolation occurs between
the LFL ($M^*\propto a+ bT^2$) and NFL ($M^*\propto T^{-2/3}$)
regimes \cite{prep,ckhz}
\begin{equation}M^*_N(y)\approx c_0\frac{1+c_1y^2}{1+c_2y^{8/3}}.
\label{UN2}
\end{equation}
Here $a$ and $b$ are constants, $c_0=(1+c_2)/(1+c_1)$, $c_1$ and
$c_2$ are fitting parameters, approximating the Landau amplitude.
Note, that our interpolative solution \eqref{UN2} is valid at low
magnetic fields, where spin dependence in Landau amplitude and
single particle spectrum is not pronounced. At high fields, when
this dependence is strong and we have full subbands spin
polarization, this interpolative solution is no more valid and we
should explicitly solve eq. \eqref{HC3} with respect to \eqref{HC2}
and \eqref{epta}. It can be shown that magnetic field $B$ enters
Landau equation only in combination $B\mu_B/T$ making $T_{\rm
max}\propto B\mu_B$ \cite{ckhz,prep}. We conclude that under the
application of magnetic field the variable
\begin{equation}\label{YTB}
y=T/T_{\rm max}\propto \frac{T}{\mu_B(B-B_{c0})}
\end{equation}
remains the same and the normalized effective mass is again
governed by eq. \eqref{UN2}. Here $B_{c0}$ is the critical magnetic
field driving both HF compound to its magnetic field tuned QCP and
corresponding N\'eel temperature to $T=0$. In some cases
$B_{c0}=0$. For example, the HF compound $\rm CeRu_2Si_2$ has
$B_{c0}=0$ and shows neither magnetic ordering nor
superconductivity \cite{takah}. In our simple model $B_{c0}$ is
taken as a parameter. In what follows, we compute the effective
mass using eq. \eqref{HC3} and employ eq. \eqref{UN2} for
qualitative analysis when considering the system at low magnetic
fields.

\begin{figure}[!ht]
\begin{center}
\includegraphics [width=0.48\textwidth]{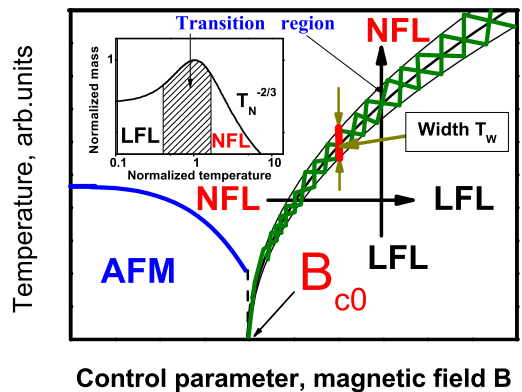}
\vspace*{-1.0cm}
\end{center}
\caption{Schematic phase diagram of $\rm YbRh_2Si_2$ \eqref{UN2}
with magnetic field as a control parameter. The vertical and
horizontal arrows show LFL-NFL and NFL-LFL transitions at fixed $B$
and $T$ respectively. At $B <B_{c0}$ the system is in AFM state.
The width of the transition region $T_w\propto T$ is shown by the
segment between two vertical arrows. Inset shows a schematic plot
of the normalized effective mass versus the normalized temperature.
Transition region, where $M^*_N$ reaches its maximum at $T/T_{\rm
max}=1$, is shown by the hatched area both in the main panel and in
the inset.}\label{MT}
\end{figure}

Now we have everything to construct the schematic phase diagram of
the HF metal $\rm YbRh_2Si_2$ at $B\ll B^*$. The phase diagram is
reported in fig. \ref{MT}. The magnetic field $B$ plays a role of
the control parameter, driving the system towards its QCP. In our
case this QCP is of FCQPT type. The FCQPT peculiarity occurs at
$B=B_{c0}$, yielding new strongly degenerate state at $B<B_{c0}$.
To lift this degeneracy, the system forms either superconducting
(SC) or magnetically ordered (ferromagnetic (FM), antiferromagnetic
(AFM) etc) states \cite{prep}. In the case of $\rm YbRh_2Si_2$,
this state is AFM one \cite{steg1}. As it follows from eqs.
\eqref{UN2} and \eqref{YTB} and seen from fig. \ref{MT}, at $B\geq
B_{c0}$ the system is in either NFL or LFL states. At fixed
temperatures the increase of $B$ drives the system along the
horizontal arrow from NFL state to LFL one. On the contrary, at
fixed magnetic field and raising temperatures the system transits
along the vertical arrow from LFL state to NFL one. The inset to
fig. \ref{MT} demonstrates the behavior of the normalized effective
mass $M^*_N=M^*/M^*_{\rm max}$ versus normalized temperature
$y=T/T_{\rm max}$ following from eq. \eqref{UN2}. The $T^{-2/3}$
regime is marked as NFL one since (contrary to LFL case where the
effective mass is constant) the effective mass depends strongly on
temperature. It is seen that temperature region $y\sim 1$ signifies
a transition regime between the LFL behavior with almost constant
effective mass and NFL one, given by $T^{-2/3}$ dependence. Thus,
temperatures $T\simeq T_{\rm max}$, shown by arrows in the inset
and main panel, can be regarded as a transition regime between LFL
and NFL states. It is seen from eq. \eqref{YTB} that the width of
the transition regime $T_w\propto T$ is proportional to
$(B-B_{c0})$. It is shown by the segment between two vertical
arrows in fig \ref{MT}. These theoretical results are in good
agreement with the experimental facts \cite{steg1,pnas}.

Our calculations of the normalized effective mass $M^*_N(T_N)$ at
fixed high magnetic field $B^*$ are shown by the solid line in the
main panel of fig. \ref{MRM}.  We recollect that in this case the
quasiparticles spins are completely polarized. This reveals the
above scaling behavior of the normalized experimental curves in
terms of the normalized variable $y=T/T_{\rm max}(B)$. It is seen
from fig. \ref{MRM} that our calculations deliver a good
description of the experiment \cite {steg1}. Namely, at elevated
temperatures ($y\simeq 1$) the LFL state first converts into the
transition one and then disrupts into the NFL state.

\begin{figure} [! ht]
\begin{center}
\vspace*{-0.8cm}
\includegraphics [width=0.49\textwidth]{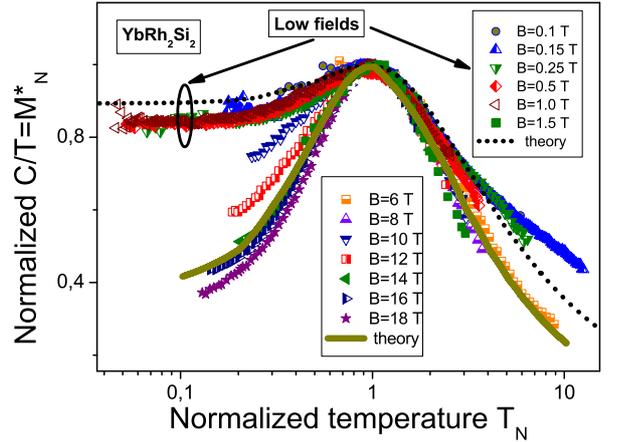}
\end{center}
\vspace*{-0.8cm} \caption{Joint behavior of the normalized
effective mass $M^*_N$ at low (upper box symbols) and high (lower
box symbols) magnetic fields extracted from the specific heat
($C/T$) measurements of the $\rm YbRh_2Si_2$ \cite{oesb}. Our
low-field calculations are depicted by the dotted line tracing the
scaling behavior of $M^*_N$. Our high-field calculations (solid
line) are taken at $B\sim B^*$ when the quasiparticle band becomes
fully polarized.}\label{f2}
\end{figure}

To perceive further the behavior of the system at high magnetic
fields, in fig. \ref{f2} we collect the curves $M^*_N(T_N)$ both at
low (symbols in the upper box in fig. \ref{f2}) and high (symbols
in the lower box) magnetic fields $B$. All curves have been
extracted from the experimental facts \cite{steg1,oesb}. It is seen
that while at low fields the low-temperature ends ($T_N \sim 0.1$)
of the curves completely merge, at high fields this is not the
case. Moreover, the low-temperature asymptotic value of $C/T=M^*_N$
at low fields is around two times more then that at high fields.
The physical reason for low-field curves merging is that the
effective mass does not depend on spin variable so that the
polarizations of subbands with $\sigma_{\uparrow}$ and
$\sigma_{\downarrow}$ are almost equal to each other. This is
reflected in our calculations, based on eq. \eqref{UN2} for low
magnetic fields $B<<B^*$. The result is shown by the dotted line in
fig. \ref{f2}.

It is also seen from fig. \ref{f2} that all low-temperature
differences between high- and low field behavior of the normalized
effective mass disappear at high temperatures. In other words,
while at low temperatures the values of $M^*_N$ for low fields are
two times more then those for high fields, at temperatures
$T_N\geq1$ this difference disappear. It is seen that these high
temperatures lie about the transition region, marked by hatched
area in the inset to fig. \ref{MT}. This means that two states (LFL
and NFL) separated by the transition region are clearly seen in
fig. \ref{f2} displaying good agreement between our calculations
(dotted line for low fields and thick line at high fields) and the
experimental points (symbols).

It is seen from fig. \ref{f2}, that at high fields $B\sim B^*$,
(symbols in the lower box) the curves $M^*_N(T_N)$ do not merge in
the low temperature LFL state. Moreover, their values decrease as
$B$ grows representing the full spin polarization of the HF band at
the highest reached magnetic fields. This behavior is opposite to
that at low fields. The corresponding theoretical curve has been
generated from the explicit numerical solution of eq. \eqref{HC3}
with respect to eqs. \eqref{HC2} and \eqref{epta}. As we have
mentioned above, at temperature raising all effects of spin
polarization smear down, yielding the restoration of NFL behavior at
$T\simeq \mu_BB$. Our high-field calculations (solid line in fig.
\ref{f2}) reflect the latter fact and are also in good agreement
with experimental facts. In order not to overload fig. \ref{f2} with
unnecessary details, we show the calculations only for the case of
the complete spin polarization.

\begin{figure} [! ht]
\begin{center}
\vspace*{-0.4cm}
\includegraphics [width=0.49\textwidth]{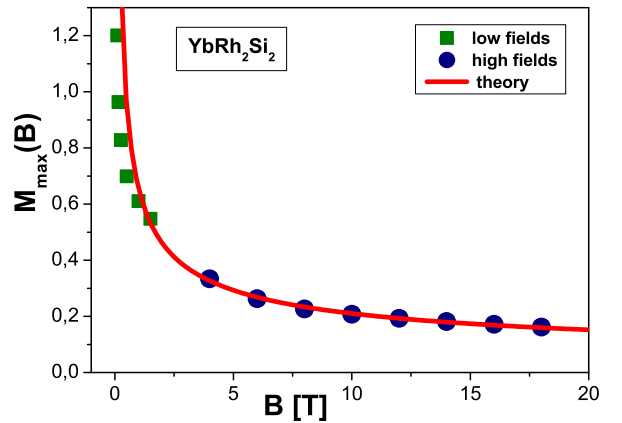}
\end{center}
\vspace*{-0.8cm} \caption{The maxima $M^*_{\rm max}(B)$ of the
functions $C/T$ versus magnetic field $B$ for $\rm YbRh_2Si_2$. The
points for low \cite{oesb} (squares) and high fields \cite{steg1}
(circles) are shown in the legend. The solid curve is approximated
by $M^*_ {\rm max}(B)\propto d/\sqrt{B-B_{c0}}$, $d$ is a fitting
parameter.}\label{MB}
\end{figure}

Figure \ref{MB} reports the maxima $M^*_{\rm max}(B)$ of the
functions in the inset to fig. \ref{MRM} versus $B$.  The solid
line represents our approximation for these maxima $M^*_{\rm
max}(B)\propto 1/\sqrt{B-B_{c0}}$ calculated within the framework
of FCQPT theory \cite{shag4,prep}. It is seen that our calculations
are in good agreement with the experimental facts in the entire
magnetic field domain. Such good coincidence indicates that at
$T\simeq \mu_BB$ the transition regime occurs and the NFL behavior
restores at high temperatures $T\sim 20$ K.

\begin{figure} [! ht]
\begin{center}
\vspace*{-0.5cm}
\includegraphics [width=0.47\textwidth]{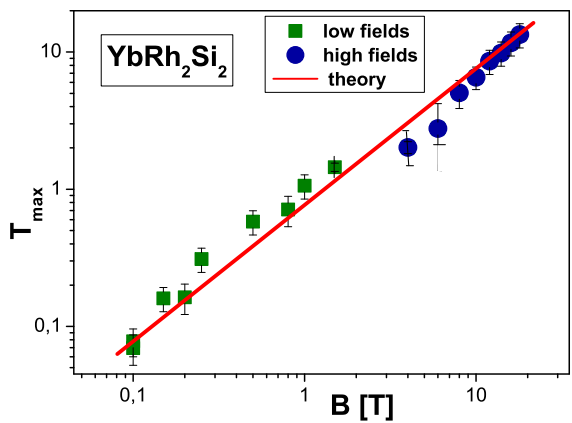}
\end {center}
\vspace*{-0.8cm} \caption{The temperatures $T_ {\rm max} (B)$ at
which the maxima of $C/T$ in $\rm YbRh_2Si_2$ (inset to fig.
\ref{MRM}) are located. Squares correspond to low-field case
\cite{oesb} and circles to high fields one \cite{steg1}.  The solid
line represents the function $T_{\rm max}\propto b(B-B_{c0})$, $b$
is a fitting parameter, see eq. \eqref{YTB}.} \label{TB}
\end{figure}

In fig. \ref{TB}, the solid squares and circles denote temperatures
$T_{\rm max}$ at which the maxima of $C/T$ (from the inset to fig.
\ref{MRM}) occur. To fit the experimental data \cite{steg1,oesb} the
function $T_ {\rm max}(B)=b(B-B_{c0})$ defined by eq. \eqref{YTB}
with $b\simeq 0.74\, {\rm K/T}$ is used. It is seen from fig.
\ref{TB} that our calculations (solid line) are in accord with
experimental facts, and we conclude that the transition regime of
$\rm YbRh_2Si_2$ is restored at temperatures $T\simeq \mu_B B$.

Consider now the magnetization $M(B,T)$ as a function of magnetic
field $B$ at fixed temperature $T$
\begin{equation}\label{CHIB} M(B,T)=\int_0^B \chi(z,T)dz,
\end{equation}
where the magnetic susceptibility $\chi$ is given by \cite{land}
\begin{equation}\label{CHI}
\chi(B,T)=\frac{\beta M^*(B,T)}{1+F_0^a}.
\end{equation}
Here, $\beta$ is a constant and $F_0^a$ is the spin-antisymmetric
Landau amplitude taken at $L=0$.

\begin{figure} [! ht]
\begin{center}
\vspace*{-0.8cm}
\includegraphics [width=0.49\textwidth]{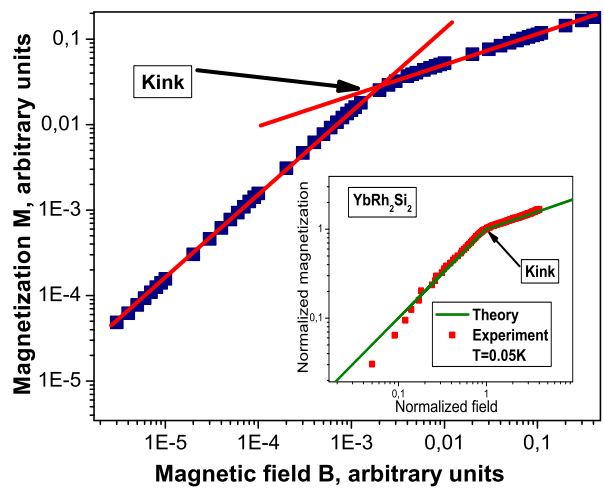}
\end{center}
\vspace*{-0.8cm} \caption{The calculated magnetization $M(B)$
(symbols) and straight lines which are guides for eye. The
intersection of the straight lines visualize the kink at the
crossover region in the fig. \ref{MT}. The inset: The field
dependencies of the normalized magnetization $M$ of $\rm{
YbRu_2Si_2}$ \cite{oesbs} at $T=0.05$ K. The kink (shown by the
arrow) is clearly seen at the normalized field $B_N=B/B_k\simeq 1$.
The solid curve represents our calculations.}\label{kink}
\end{figure}

Our calculations show that the magnetization exhibits a kink at
some magnetic field $B=B_k$. The experimental magnetization
demonstrates the same behavior \cite{steg,oesbs}. We use $B_k$ and
$M(B_k)$ to normalize $B$ and $M$ respectively. In the normalized
variables, there are no coefficients $\beta$ and $(1+F_0^a)$ so
that $\chi\propto M^*$ \cite{prep} and we can once more use eq.
\eqref{HC3} to calculate the magnetic susceptibility $\chi$. The
normalized magnetization $M(B)/M(B_k)$ both extracted from
experiment (symbols) and calculated one (solid line), are reported
in the inset to fig. \ref{kink}. It shows that our calculations are
in good agreement with the experiment. All the data exhibit the
kink (shown by the arrow) at $B_N\simeq 1$ taking place as soon as
the system enters the transition region. This region corresponds to
the magnetic fields where the horizontal arrow in fig. \ref{MT}
crosses the hatched area. To illuminate the kink position, in the
fig. \ref{kink} we present the $M(B)$ dependence in logarithmic -
logarithmic scale. In that case the straight lines show clearly the
change of the slope (power in logarithmic scale) of $M(B)$ at the
kink.

\begin{figure} [! ht]
\begin{center}
\vspace*{-0.5cm}
\includegraphics [width=0.47\textwidth]{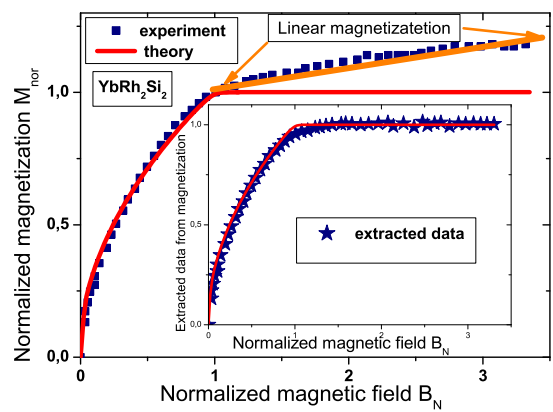}
\end {center}\vspace{-0.8cm}
\caption {The normalized magnetization $M_{\rm nor}$ as a function
of the normalized magnetic field $B_N$. Line represents our
calculations and squares represent experimental points
\cite{steg1}. The linear dependence $M_{\rm nor}(B_N)$ is marked by
the arrows. The inset demonstrates the experimental data (stars)
with the subtracted high-field linear part. Our calculations is
shown by the solid line.}\label{kink2}
\end{figure}

At magnetic field $B\simeq B^*$ the quasiparticle band becomes
fully polarized and a new kink appears \cite{steg1,mal}. We call
this kink as the second one. Our calculations of the normalized
magnetization (line) and the experimental points (squares) are
shown in fig. \ref{kink2}. In that case both the magnetization and
the field are normalized by the corresponding values at the second
kink position.

In the fig. \ref{kink2}, we plot our theoretical normalized (in the
second kink point) magnetization along with experimental one. Good
coincidence is seen everywhere except the high-field part at
$B_N\geq 1$. Here, the experimental normalized magnetization
$M_{\rm nor}$ exhibits a linear dependence on $B_N$ (marked by two
arrows), while the calculated magnetization is approximately
constant.  Such a behavior is the intrinsic shortcoming of the HF
liquid model that accounts for only heavy electrons and omits the
conduction electrons of other kind \cite{c1,c2}. Thus, we can
consider the high-field (at $B_N>1$) part of the magnetization as
the contribution which is not included in our theory. To separate
this contribution from the experimental magnetization curve, we
(numerically) differentiate it, then subtract constant part at
$B_N>1$ and integrate back the resulting curve. The coincidence
between our calculations depicted by the solid curve and processed
experimental data shown by the stars is reported in the inset to
fig. \ref{kink2}. As we can see now, the coincidence between the
theory and experiment is good in the entire magnetic field domain.
Taking into account the obtained results displayed in figs.
\ref{f2}, \ref{MB}, \ref{TB}, \ref{kink} and \ref{kink2}, we
conclude that the HF system of $\rm{ YbRu_2Si_2}$ evolves
continuously under the application of magnetic field. This fact is
in agreement with experimental observations \cite{prl}.

To summarize, here we have analyzed the thermodynamic properties of
$\rm YbRh_2Si_2$ at both low and high magnetic fields. Our
calculations allow us to conclude that in magnetic field the HF
system of $\rm YbRh_2Si_2$ evolves continuously without a
metamagnetic transition and possible localization of heavy $4f$
electrons. Under the application of magnetic field at low
temperatures, the HF system demonstrates the LFL behavior, while at
elevated temperatures the system enters the transition region
followed by the NFL behavior. Our calculations are in good
agreement with experimental facts in the entire temperature and
magnetic field domains under consideration.

This work was supported in part by the RFBR grant No. 09-02-00056.

\end{document}